\documentclass[pra,aps,twocolumn,10pt,showpacs,groupedaddress,superscriptaddress,floatfix]{revtex4-2}
\usepackage{amsmath,amsfonts,amssymb,graphics,graphicx,epsfig,color,times}
\usepackage[utf8]{inputenc} 
\usepackage{color}
\usepackage{bbm, dsfont} 
\usepackage{subfigure}
\usepackage{hyperref}
\usepackage{mathrsfs}
\usepackage{verbatim}
\begin{document}
\title{Resonant Dipole-Dipole Interactions in Electromagnetically Induced Transparency}
\author{H. H. Jen}
\email{sappyjen@gmail.com}
\affiliation{Institute of Atomic and Molecular Sciences, Academia Sinica, Taipei 10617, Taiwan}
\affiliation{Physics Division, National Center for Theoretical Sciences, Taipei 10617, Taiwan}

\author{G.-D. Lin}
\affiliation{Department of Physics, National Taiwan University, Taipei 10617, Taiwan}
\affiliation{Physics Division, National Center for Theoretical Sciences, Taipei 10617, Taiwan}

\author{Y.-C. Chen}
\affiliation{Institute of Atomic and Molecular Sciences, Academia Sinica, Taipei 10617, Taiwan}
\affiliation{Center for Quantum Technology, Hsinchu 30013, Taiwan}
\affiliation{Physics Division, National Center for Theoretical Sciences, Taipei 10617, Taiwan}

\date{\today}
\renewcommand{\k}{\mathbf{k}}
\renewcommand{\r}{\mathbf{r}}
\newcommand{\parallelsum}{\mathbin{\!/\mkern-5mu/\!}}
\def\p{\mathbf{p}}
\def\R{\mathbf{R}}
\def\bea{\begin{eqnarray}}
\def\eea{\end{eqnarray}}
\def\bee{\begin{equation}}
\def\eee{\end{equation}}
\begin{abstract}
Resonant dipole-dipole interaction (RDDI) emerges in strong light-matter interacting systems, which leads to many fascinating phenomena like cooperative light scattering and collective radiations. Here we theoretically investigate the role of RDDI in electromagnetically induced transparency (EIT). The resonant dipole-dipole interactions manifest in the cooperative spontaneous emission of the probe light transition, which give rise a broadened linewidth and associated collective frequency shift. This cooperative linewidth originates from the nonlocal and long-range RDDI, which can be determined by the atomic density, optical depth, and macroscopic length scales of the atomic ensemble. We present that EIT spectroscopy essentially demonstrates all-order multiple scattering of RDDI. Furthermore, we find that EIT transparency window becomes narrower as the cooperative linewidth increases, which essentially reduces the storage efficiency of slow light as EIT-based quantum memory application.
\end{abstract}
\maketitle
\section{Introduction}

Electromagnetically induced transparency (EIT) \cite{Harris1990, Harris1997, Lukin2003, Fleischhauer2005} has been a robust technique to store light field while preserving its quantum coherence with little dissipations.\ This facilitates quantum interface \cite{Hammerer2010} of light and matter with high efficiency and controllability. Due to the dramatic modification of the dispersive properties accompanying with the induced transparency, the probe field propagates through the interacting medium with a reduced group velocity \cite{Hau1999, Schnorrberger2009}.\ Furthermore, the picture of dark state polariton \cite{Fleischhauer2002} can be adopted to describe light-matter dynamics, where light can be stopped and transferred to atomic coherences, and then retrieved back to light again by turning off and on the control field. In addition to neutral atoms \cite{Lukin2003, Fleischhauer2005, Hsiao2018, Wang2019}, EIT has been implemented in various platforms including quantum degenerate gases \cite{Hau1999, Schnorrberger2009}, Rydberg atoms \cite{Saffman2010, Pritchard2010, Dudin2012, Peyronel2012}, single atoms in a cavity \cite{Mucke2010, Kampschulte2014}, embedded Fe nuclei \cite{Rohlsberger2012}, color centers in a diamond \cite{Hemmer2001, Acosta2013}, semiconductor quantum well \cite{Serapiglia2000}, and crystals doped with rare-earth ions \cite{Ham1997, Turukhin2001, Longdell2005,Baldit2010, Heinze2013}.

For the conventional EIT theory \cite{Lukin2003, Fleischhauer2005}, it is mainly based on the classical electrodynamics with a mean-field treatment of dipole operators that govern the light-matter interactions \cite{QO:Scully}.\ This theory has be extended to investigate a strongly correlated quantum degenerate gas \cite{Jiang2009}, where intriguing quantum many-body effects emerge in EIT spectroscopy when the atoms are coupled to the low-lying Rydberg states \cite{Jen2013, Jen2014}. On the other hand, even when the atomic ground state is not in a quantum degenerate regime, which is the case for cold atoms in general, the conventional EIT theory neglects a full account of the effect of resonant dipole-dipole interactions (RDDIs) \cite{Lehmberg1970} in the dissipation channels of both probe and control fields. This effect is crucial especially for atoms of a high density or with a large optical depth (OD), which leads to superradiance \cite{Dicke1954, Gross1982} and other cooperative spontaneous emissions. Therefore, the conventional EIT theory has to be extended in this regard to include the essential collective scattering events mediated by the nonlocal and long-range RDDIs.  

There has been a huge progress in recent experiments which demonstrate the effect of RDDIs in the collective radiations. These rescattering emissions between every two atoms can enhance the light emission in a dense atomic medium \cite{Bromley2016}, and are responsible for subradiant emissions in neutral atoms \cite{Guerin2016}, plasmonic nanocavities \cite{Sonnefraud2010}, ultracold molecules \cite{McGuyer2015}, and metamolecules \cite{Jenkins2017}. The associated collective frequency shifts can also be observed in various two-level atomic systems of the embedded Fe nuclei \cite{Rohlsberger2010}, atomic vapor layer \cite{Keaveney2012}, ions \cite{Meir2014}, and cold atoms \cite{Pellegrino2014, Jennewein2016, Jenkins2016, Roof2016}. However, this evident cooperative phenomenon has been elusive in the EIT platform of $\Lambda$-type atomic configuration as shown in Fig. \ref{fig1_EIT}. Only recently a semiclassical treatment of light-matter interactions is applied in studying EIT properties with comparable probe and control fields \cite{Oliveira2021}, where light-mediated interactions can modify the transparency window. Here in this paper on the contrary, we include the RDDI in the conventional EIT theory involving up to single atomic excitation and derive the effective linewidth broadening and frequency shift in EIT transmission. We show that the probe field propagates through a narrowing EIT transmission window when RDDI is significant. This reflects the nonlocal nature of RDDI, which incorporates all pairwise dipole-dipole interactions and can be observable in the future EIT measurements. 

The paper is organized as follows. In Sec. II, we consider a EIT scheme in an atomic ensemble, where we obtain the transmission property from the coupled equations with RDDI and their steady-state solutions. We next investigate the linewidth of the probe light transition in Sec. III and present the results of multiple scattering of RDDI in the transmission spectrum in Sec. IV. We also discuss the extension beyond the local field approximation in Sec. V. Finally we discuss and conclude in Sec. VI. In Appendix A, we obtain Maxwell-Bloch equations with RDDI in the probe field transition, and in Appendix B, we present the steady-state solutions from Maxwell-Bloch equations. 

\section{RDDI effect in EIT}

\subsection{Hamiltonian and Lindblad forms}

\begin{figure}[t]
\centering
\includegraphics[width=8.5cm,height=6.4cm]{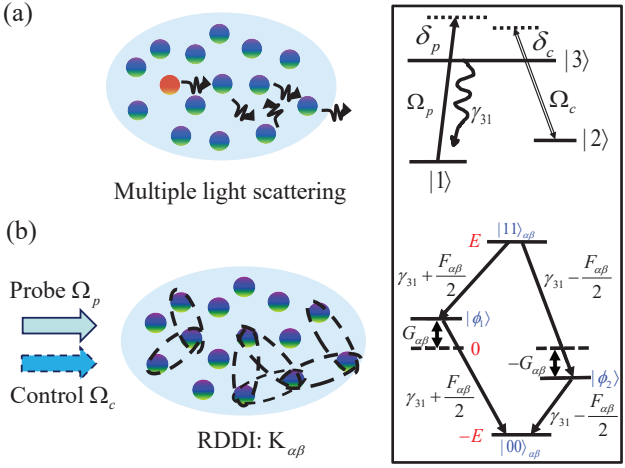}
\caption{Schematic plot of multiple scattering of light and EIT scheme with RDDI. (a) An excited atom emits and scatters light through many rescattering events between atoms before leaving the media and being observed. (b) A conventional EIT scheme with control $\Omega_c$ and probe fields $\Omega_p$ interacting with three-level configurations in the upper inset plot with spontaneous emission decay rate $\gamma_{31}$. Probe and control field detunings are denoted as $\delta_p$ and $\delta_c$, respectively. With the effect of pairwise RDDI $K_{\alpha\beta}$ which involves a collective decay $F_{\alpha\beta}$ and frequency shift $G_{\alpha\beta}$ between every other atoms, a level diagram for two atoms is presented in the lower inset plot as a demonstration, where super- and subradiant eigen decay rates $\gamma_{31}\pm F_{\alpha\beta}/2$ are associated with eigenstates $|\phi_{1(2)}\rangle=(|01\rangle_{\alpha\beta}\pm|10\rangle_{\alpha\beta})/\sqrt{2}$, respectively.}\label{fig1_EIT}
\end{figure}

The effect of RDDI in light propagation is mainly investigated in $N$ two-level atoms, where a coupled-dipole model along with Maxwell-Bloch equation can describe light-atom dynamics \cite{Jennewein2018}. This treatment is valid when a weak driving field is considered, which effectively takes into account of the effects from both RDDI and the propagating light field in the time-evolving atomic coherences. This leads to an effective linewidth broadening and associated frequency shift for the light field transition \cite{Roof2016, Araujo2016, Sutherland2016}. We follow this similar treatment in our EIT setup using $\Lambda$-type atoms in Fig. \ref{fig1_EIT}, where a probe field $\Omega_p$ couples the ground $|1\rangle$ to the excited state $|3\rangle$ while a control field $\Omega_c$ couples $|3\rangle$ to the other hyperfine ground state $|2\rangle$. The ground state coherence establishes on absorbing a probe and emitting a control photon, which excites one of $N$ atoms to $|2\rangle$ collectively. Therefore, a probe photon exchanging with atomic coherence forms a dark-state polariton which propagates through the medium, almost immune to the intrinsic spontaneous emission $\gamma_{31}$, and thus keeps its waveform intact inside the medium. Since light couples the atoms collectively, RDDI in the dissipation process should have an effect on EIT spectroscopy, especially for a dense atomic cloud. 

For a conventional EIT setup \cite{Hsiao2018} as an example, the system parameters are $N\sim 5\times 10^9$ in a cloud size of $3\times 3\times 14$ mm$^3$, which has an average atomic density $\sim 4\times 10^{10}$ cm$^{-3}$. For a peak density around $10^{11}$ cm$^{-3}$ and considering an interaction volume determined by probe field propagation along the long axis with a cylindrical geometry of $\pi(0.2)^2\times 14$ mm$^3$, we have $N\sim 1.8\times 10^8$ with an optical density $\sim 400$. This parameter regime is particularly relevant in conventional EIT experiments as quantum memory applications, where the propagation effect of the probe field becomes essential. In this parameter regime, there is no possible numerical simulation using present computation technology. Therefore, we intend to provide an analytical calculation of EIT spectroscopy in a cigar shape of atoms with a relatively large optical density under a moderate atomic density in the order of $10^{10}$ to $10^{11}$ cm$^{-3}$. In such a long sample of atoms, we note that the Beer-Lambert law for light propagation can be retrieved and sustained, which would otherwise break down in a thin slab geometry of dense atoms \cite{Chomaz2012}.  

Here we include and focus on the effect of RDDI in the probe transition and investigate how it modifies the conventional EIT theory. This RDDI should also manifest in the control field transition and between two ground states, with respective intrinsic decay rates $\gamma_{32}$ and $\gamma_{21}$. However, a relatively large $\Omega_c$ and vanishing populations in $|2\rangle$ and $|3\rangle$ are legitimate to neglect this effect on $\gamma_{32}$, in contrast to the weak $\Omega_p$. The ground state decoherence also involves an inappreciable dipole transition, and therefore the effect of RDDI on $\gamma_{21}$ should be insignificant compared to $\gamma_{31}$. 

The Hamiltonian for EIT in interaction picture reads
\bea
\hat H_I =&& \hbar\delta_p \sum_{\mu=1}^N \hat\sigma_{11}^{\mu} + \hbar\delta_c \sum_{\mu=1}^N \hat\sigma_{22}^\mu \nonumber\\
&& + \sum_{\mu=1}^N \left(-\frac{\hbar\Omega_p}{2}e^{i\k_p\cdot \r_\mu}\hat\sigma_{13}^{\mu,\dag} + h.c.\right)\nonumber\\
&& + \sum_{\mu=1}^N \left(-\frac{\hbar\Omega_c}{2}e^{i\k_c\cdot \r_\mu}\hat\sigma_{23}^{\mu,\dag} + h.c.\right),\label{H_I}
\eea
where the wave vectors of the fields are $\k_p$ and $\k_c$ with the Rabi frequencies $\Omega_p$ and $\Omega_c$ respectively, and the atomic dipole operators are $\hat\sigma_{ij}^\beta$ $\equiv$ $|i\rangle_\beta\langle j|$, $\hat\sigma_{ij}^{\beta,\dag}\equiv\hat\sigma_{ji}^\beta$. The detunings are defined as $\delta_p$ $=$ $\omega_p$ $-$ $\omega_{31}$ and $\delta_c$ $=$ $\omega_c$ $-$ $\omega_{32}$, where the central frequencies of light fields are $\omega_p$, $\omega_c$, and the atomic transition frequencies are $\omega_{31}$, $\omega_{32}$. To account for RDDI effect in the probe field transition, we use Lindblad forms \cite{Tannoudji1992} to fully describe the system dynamics, and the Master equation for arbitrary operators $\hat Q$ in Heisenberg picture becomes 
\bea
\frac{d\hat Q}{dt}=&&-\frac{i}{\hbar}[\hat Q, \hat H_I]-i\sum_{\mu\neq\nu,\nu}^N[\hat Q, G_{\mu\nu}\hat\sigma_{31}^\mu\hat\sigma_{13}^\nu]\nonumber\\
&& +\mathcal{L}_p[\hat Q]+\mathcal{L}_c[\hat Q]+\mathcal{L}_g[\hat Q],\label{Q}
\eea
where 
\bea
\mathcal{L}_p[\hat Q]\equiv&&-\sum_{\mu,\nu}^N\frac{F_{\mu\nu}}{2}(\hat\sigma_{31}^\mu\hat{\sigma}_{13}^\nu\hat Q+\hat Q\hat\sigma_{31}^\mu\hat{\sigma}_{13}^\nu-2\hat\sigma_{31}^\mu \hat Q\hat{\sigma}_{13}^\nu),\nonumber\\\\
\mathcal{L}_c[\hat Q]\equiv&&-\sum_{\mu,\nu}^N\gamma_{32}(\hat\sigma_{32}^\mu\hat{\sigma}_{23}^\nu\hat Q+\hat Q\hat\sigma_{32}^\mu\hat{\sigma}_{23}^\nu-2\hat\sigma_{32}^\mu \hat Q\hat{\sigma}_{23}^\nu),\nonumber\\\\
\mathcal{L}_g[\hat Q]\equiv&&-\sum_{\mu,\nu}^N\gamma_{21}(\hat\sigma_{21}^\mu\hat{\sigma}_{12}^\nu\hat Q+\hat Q\hat\sigma_{21}^\mu\hat{\sigma}_{12}^\nu-2\hat\sigma_{21}^\mu \hat Q\hat{\sigma}_{12}^\nu).\nonumber\\
\eea

The cooperative spontaneous decay rates $F_{\mu\nu}$ and frequency shifts $G_{\mu\nu}$ can be expressed as \cite{Lehmberg1970}
\bea
F_{\mu\nu}(\xi)\equiv&&
\frac{3\Gamma}{2}\bigg\{\left[1-(\hat{d}\cdot\hat{r}_{\mu\nu})^2\right]
\frac{\sin\xi}{\xi}\nonumber\\
&&+\left[1-3(\hat{d}\cdot\hat{r}_{\mu\nu})^2\right]
\left(\frac{\cos\xi}{\xi^2}-\frac{\sin\xi}{\xi^3}\right)\bigg\},\label{F}\\
G_{\mu\nu}(\xi)\equiv&&\frac{3\Gamma}{4}\bigg\{-\Big[1-(\hat{d}\cdot\hat{r}_{\mu\nu})^2\Big]\frac{\cos\xi}{\xi}\nonumber\\
&&+\Big[1-3(\hat{d}\cdot\hat{r}_{\mu\nu})^2\Big]
\left(\frac{\sin\xi}{\xi^2}+\frac{\cos\xi}{\xi^3}\right)\bigg\}\label{G}, 
\eea
where $\Gamma$ $=$ $2\gamma_{31}$ $\equiv$ $\omega_{31}^3d^2/(3\pi\epsilon_0\hbar c^3)$, is the intrinsic decay rate, and $\hat d$ is the dipole orientation with dipole moment $d$. The dimensionless scale of mutual separation is $\xi$ $=$ $|\k_p| r_{\mu\nu}$ with $r_{\mu\nu}$ $=$ $|\r_\mu-\r_\nu|$. The above indicates the long-range nature of dipole-dipole interactions, which is responsible for the cooperative radiations of superradiance or subradiance. As $\xi$ $\rightarrow$ $0$, $F_{\mu\nu}$ approaches $\Gamma$ while $G_{\mu\nu}$ becomes divergent, showing an incomplete quantum optical treatment in this limit.\ For $\xi$ $\gg$ $2\pi$ or $r_{\mu\nu}$ $\gg$ $\lambda$ (transition wavelength), both cooperative decay rates and frequency shifts diminish, which reaches the noninteracting regime of independent atoms.

\subsection{Maxwell-Bloch equations and transmission coefficient}

From the Maxwell-Bloch equations we obtain in Appendix A, we have truncated the coupled equations at the first-order cumulants \cite{Kubo1962} or one-body expectation values. This allows self-consistent and dynamically-coupled equations, where a hierarchy of many-body correlations are assumed insignificant. We define the cross-grained and slow-varying atomic coherence in the probe transition as $\tilde{\sigma}_{13}=N_z^{-1}\sum_{\beta=1}^{N_z}\hat\sigma_{13}^\beta(z,t) e^{i\omega_pt-i\k_p\cdot\r_\beta}$, and from Eq. (\ref{sigma13_app}) we obtain 
\bea
\frac{d}{d\tau}\tilde{\sigma}_{13} &\approx& (i\delta _{p}-\gamma_{31})\tilde{\sigma}_{13}+i\frac{\Omega _{c}}{2}\tilde{\sigma}_{12}+i\frac{\Omega_p}{2}\nonumber\\
&&-\frac{1}{N_z}\sum_{\alpha=1}^{N_z}\sum_{\beta\neq\alpha}^NK_{\alpha\beta}\tilde\sigma_{13}^\beta,\label{sigma13_m}
\eea 
where $\tau$ $=$ $t$ $-$ $z/c$ in a co-propagating frame, slow-varying atomic operator $\tilde{\sigma}_{13}^\beta\equiv \hat\sigma_{13}^\beta e^{i\omega_pt-i\k_p\cdot\r_\beta}$, and $K_{\alpha\beta}$ $\equiv$ $(F_{\alpha\beta}$ $+$ $i2G_{\alpha\beta})e^{-i\k_p\cdot\r_{\alpha\beta}}/2$. In Eq. (\ref{sigma13_m}), we have assumed that most of the atoms are in the ground state $\tilde\sigma_{11}$$\approx$$1$, which is valid in the linear response regime with a weak probe field. The $N_z\equiv N/M$ denotes the number of atoms in $M$ cross-grained sections along the propagation direction in $\hat z$, which is introduced as functional and does not come into play in continuous limit of field quantization under large $N$, $N_z$, and $M$ \cite{Drummond1987}.  Equation (\ref{sigma13_m}) shows the nonlocal ($r_{\alpha\beta}^{-1}$ in long range) and linear couplings between atoms throughout the medium via RDDI, in contrast to Rydberg van der Waals interactions ($\propto r_{\alpha\beta}^{-6}$) \cite{Pritchard2010, Sevincli2011} where nonlinear interactions result from two or more Rydberg excitations, and a dipole-blockade sphere can be established due to large energy shift.

We then solve the steady-state ground state coherence as shown in Appendix A,
\begin{equation}
\tilde \sigma_{12}=\frac{-\Omega_c^*/2}{\delta_2+i\gamma_{21}}\tilde\sigma_{13},\label{sigma12_steady_m}
\end{equation}
where $\delta_2=\delta_p-\delta_c$ and $\delta_c=\omega_c-\omega_{32}$ are two-photon and control-field detunings, respectively. From Eqs. (\ref{sigma13_m}) and (\ref{sigma12_steady_m}), we iteratively solve the probe field coherence in perturbations of $K_{\alpha\beta}$ with local field approximations \cite{Sevincli2011} in Appendix B. 

Along with Maxwell-Bloch equation,
\bea
\frac{d}{d z}\Omega_p=&&\frac{iD\Gamma}{2L}\tilde{\sigma}_{13},
\eea
we obtain the EIT transmission $T=|\Omega_p(L)/\Omega_p(z=0)|^2$ up to the $M$th order of perturbations as
\bee
T_M=\exp\left[\frac{D\Gamma}{2}\textrm{Re}\left(\sum_{m=0}^M \frac{f_C^m}{A^{m+1}}\right)\right],\label{transmission_order}
\eee
where we define the optical depth as $D$ $\equiv$ $\rho\sigma L$ with an atomic density $\rho$, scattering cross section $\sigma$, and propagation length $L$. We further use the intrinsic decay rate $\Gamma$ $=$ $2\gamma_{31}$ as a universal measure with or without RDDI. The above, as $M\rightarrow\infty$, leads to 
\bee
T=\exp\left[\frac{D\Gamma}{2}\textrm{Re}\left(\frac{1}{A-f_{C}}\right)\right],\label{transmission}
\eee
where
\bea
A\equiv&& i\delta_p-\gamma_{31}-\frac{i\Omega_c^2/4}{\delta_2+i\gamma_{21}},\\
f_{C}\equiv&& \frac{1}{N_z}\sum_{\alpha=1}^{N_z}\sum_{\beta\neq\alpha}^{N}K_{\alpha\beta}.\label{fc}
\eea
Equation (\ref{transmission}) further indicates the effective collective frequency shift and linewidth, respectively, 
\bee
\tilde\delta_p=\delta_p - \textrm{Im}[f_{C}],~
\tilde\gamma_{31}=\gamma_{31}+\textrm{Re}[f_{C}],\label{LW_m}
\eee
where RDDI directly modify the transparency condition and linewidth. We note that for singly-excited collective states in $N$ two-level atoms, we typically have $N$ eigenmodes and associated eigenstates to fully describe the system dynamics \cite{Jen2016_SR, Sutherland2016}. It is the coherent forward scattering in the paraxial Maxwell-Bloch equation \cite{Sutherland2016, Jennewein2018} that makes Eq. (\ref{LW_m}) a simple relation with a modification in frequency shift and linewidth directly from the RDDI effect in the probe transition. 

The transmission of Eq. (\ref{transmission}) is valid only when $|A|\gg|f_C|$, which is easily satisfied since $\Omega_c/|\delta_2+i\gamma_{21}|\gg 1$ inside the transparency window or $|A|\approx |\delta_p|\gg |f_C|$ when way off single-photon resonance. Near strong absorption in EIT spectrum, that is, at small transmission $T$, the iterative perturbations of $K_{\alpha\beta}$ could fail since $|A|\sim\gamma_{31}$ which can be exceeded by $|f_{C}|$. However, for a large optical depth when $D$ $\gg$ $|f_{C}|/\gamma_{31}$, the transmission becomes vanishing, and therefore Eq. (\ref{transmission}) still holds in this limit.

The clear advantage of EIT over fluorescence of two-level atoms on revealing cooperative linewidth manifests in the role of control field. It is this large energy scale that validates infinite (all-order) perturbations of RDDI, and thus $f_C$ can genuinely characterize the collective frequency shift and linewidth broadening of the probe photon. On the contrary, in two-level atoms, multiple scattering of RDDI can not be quantitatively calculated unless $|f_C|\ll \gamma_{31}$, which essentially shows no cooperative effect whatsoever. To get around this issue, a direct Monte Carlo simulation can be implemented to configure atomic spatial distributions, which leads to a coupling matrix involving RDDI. Numerically diagonalizing this coupling matrix between any two radiating dipoles effectively includes all-order scattering of RDDI \cite{Guerin2016,Bromley2016}. However, this brute-force numerical method only applies well in a small Hilbert space, that is, of a small number of atoms (up to several thousands) with single \cite{Guerin2016,Bromley2016,Jen2016_SR} or few-photon excitations \cite{Jen2017_MP}. A recent attempt of numerical simulation has tackled more than $10^5$ atoms using an iterative method in calculating scattering rates \cite{Robicheaux2020} with repeated exact diagonalization of a subset of atoms. Therefore, comparing several millions of atoms operated in fluorescence or even $10^8$ atoms in EIT experiments, a huge gap nevertheless exists between theory and measurements \cite{Chomaz2012}, making direct comparison and prediction improbable in the near future. 

It seems that modeling a much smaller number of atoms is feasible in our case, say $10^5$ atoms to the most under the present computation capability. However, for the typical atomic density used in EIT experiments we consider here, a thousand-fold decrease of atoms (from $10^8$ to $10^5$) means a reduction of an optical depth below one (from around $500$ to $0.5$) if the cross-section area of atoms, determined by a tightly focused probe field, is kept the same. This indicates negligible RDDI effect on EIT properties up to $N=10^5$. The parameter regime in most of EIT experiments resides in the category of a moderate atomic density with high optical depth and therefore large number of atoms as well, and this poses a dilemma in direct numerical simulations of the regime we consider here. To reveal the RDDI effect in a small sample numerically, one can consider an even higher density of atoms, where however Maxwell-Bloch equations and Beer-Lambert law in this regime break down \cite{Chomaz2012}, and it is beyond our consideration in the paper. Meanwhile, EIT properties in a small sample with a high atomic density remain an open question and deserve future explorations.   

We note that there is a recent effort applying the concept of renormalization group method \cite{Andreoli2021} to reduce the complexity of large system under strong RDDI into an ensemble of inhomogeneously broadened and weakly interacting atoms. This is the essence of cooperativity in light-matter interactions, where many-atom physics could not be easily and simply extracted from a few-atom case. This complexity even augments when nonlocal RDDI engages in the dissipation process, where atom-atom correlations are definite to play important roles \cite{Pellegrino2014, Jennewein2016, Jenkins2016} in fluorescence measurements.      

\section{Linewidth of probe light}

We proceed to give an estimate of the effective linewidth $f_C$ in Eq. (\ref{transmission}) resulting from RDDI in $\Lambda$-type EIT scheme. $f_{C}$ is defined in Eq. (\ref{fc}) in a form of discrete sum, where $\sum_{\beta\neq\alpha}^{N}K_{\alpha\beta}$ can be treated with a referenced atom at position $\r_\alpha$. We will prove later that the leading order result does not involve where $\r_\alpha$ is, and therefore the $N_z$ cross-grained average can just be replaced by the result of single referenced atom.

We first calculate the part of cooperative linewidth $F_{\alpha\beta}$ in $f_C$, and the associated frequency shift can be derived accordingly from cooperative linewidth. The continuous form of $\tilde\gamma_{31}=\sum_{\beta=1}^{N}F_{\alpha\beta}e^{-i\k_p\cdot\r_{\alpha\beta}}/2$ including the intrinsic decay rate $\gamma_{31}$ when $\beta$ $=$ $\alpha$ becomes \cite{Lehmberg1970, Rehler1971, Mazets2007}
\bea
\tilde\gamma_{31}=&&N\gamma_{31}\int_{-\infty}^\infty dxdydz(\pi^{3/2}R_\perp^{-2}R_L^{-1})e^{-\frac{x^2+y^2}{R_\perp^2}}e^{-\frac{z^2}{R_L^2}} \nonumber\\
&&\times\int\frac{3}{8\pi}\left(1-\frac{\sin^2\theta}{2}\right)d\Omega_\k e^{-i\k_p\cdot(\r_\alpha-\r)}e^{i\k\cdot(\r_\alpha-\r)},\nonumber\\
\eea
where we have assumed Gaussian distributions with $R_\perp$ and $R_L$ respectively for the transverse and longitudinal (propagation) length scales. The $4\pi$ solid angle integration $d\Omega_\k$ in the above comes from the original integral form of $F_{\alpha\beta}$ \cite{Lehmberg1970}, where circular polarizations are assumed and $|\k|=|\k_p|$. We further assume $\r_\alpha$ lies on the $\hat x-\hat z$ plane with an angle $\theta'$ to $\hat z$, and we obtain
\bea
\tilde\gamma_{31}=&&\int\frac{3N\gamma_{31}}{8\pi}(1-\frac{\sin^2\theta}{2})\sin\theta d\theta d\phi\nonumber\\
&&\times e^{-|\k_p|^2R_\perp^2\sin^2\theta/4} e^{-|\k_p|^2R_L^2(1-\cos\theta)^2/4} \nonumber\\
&&\times e^{i|\k_p||\r_\alpha|(\cos\theta\cos\theta'+\sin\theta\cos\phi\sin\theta'-\cos\theta')}.
\eea
We further integrate out the part of $\int d\phi$, which gives $2\pi J_0(|\k_p||\r_\alpha|\sin\theta\sin\theta')$, a Bessel function of the first kind. Since exponentially small weightings for $\theta$ $\sim$ $\pi/2$ in the above Gaussian functions, we consider $\theta\gtrsim 0$ and $J_0(x')$ $\approx$ $1-x'^2/4$ for small $x'$. We then take its leading order, set $x$ $=$ $\cos\theta$, and we obtain
\bea
\tilde\gamma_{31}=&&\frac{3N\gamma_{31}}{8}\int_{-1}^1 (1+x^2)dx e^{-\frac{|\k_p|^2R_\perp^2(1-x^2)}{4}} e^{-\frac{|\k_p|^2R_L^2(1-x)^2}{4}}\nonumber\\
&&\times e^{-i|\k_p||\r_\alpha|\cos\theta'(1-x)}.\label{int_r}
\eea
When $R_\perp$$=$$R_L$ and with $|\k_p|R_L\gg 1$, we can further simplify the above to 
\bea
\tilde\gamma_{31}=&&\frac{3N\gamma_{31}}{8}\frac{2}{|\k_p|^2R_L^2/2+i|\k_p||\r_\alpha|\cos\theta'}\\
&&\approx\frac{3N\gamma_{31}}{2|\k_p|^2R_L^2}.\label{Mazet_comp}
\eea
From the above, we can express it in terms of an effective $D_c$$=$$2\rho\sigma R_L$ with $\sigma$$=$$3\lambda^2/(4\pi)$ and use the total volume $V$$=$$\pi^{3/2}R_L^3$ of Gaussian density distributions and the averaged density $\rho$$=$$N/V$, and we have $\tilde\gamma_{31}/\gamma_{31}$$=$$\sqrt{\pi}D_c/4$. We note that Eq. (\ref{Mazet_comp}) has the same dependence of sizes as in a spherical geometry with considering only long-range terms in RDDI \cite{Mazets2007}. 

In general for a cigar shape, let $a$ $=$ $|\k_p|^2R_\perp^2/4$, $b$ $=$ $|\k_p|^2R_L^2/4$, $m$ $\equiv$ $b/a$, and from Eq. (\ref{int_r}) we obtain 
\bea
\frac{\tilde\gamma_{31}}{\gamma_{31}}=&&\frac{3N}{8}\frac{1}{4a^{3/2}(m-1)^{5/2}}\bigg\{e^{\frac{a}{m-1}}\sqrt{\pi}(4am^2-4ma+m\nonumber\\
&&-1+2a)\left[\textrm{Erf}\left(\frac{(2m-1)\sqrt{a}}{\sqrt{m-1}}\right)-\textrm{Erf}\left(\frac{\sqrt{a}}{\sqrt{m-1}}\right)\right]\nonumber\\
&&+2\sqrt{a}\sqrt{m-1}(e^{-4ma}-2m+1)\bigg\},\label{general}
\eea
where Erf is the error function. We note that the above expression provides a general recipe to describe the collective decay constant in a cigar shape of atoms with Gaussian distributions, which are relevant for realistic EIT experiments on long samples of atoms \cite{Hsiao2018}. In the extreme needle-like shape where $ma$ $\gg$ $1$ and $m$ $\gg$ $1$, we can further have a simple form as 
\bea
\tilde\gamma_{31}/\gamma_{31}=\frac{3N}{8}\frac{\sqrt{\pi}4am^2}{4a^{3/2}m^{5/2}}\times\frac{1}{2}=\frac{3\sqrt{\pi}}{8}\frac{N}{|\k_p|R_L},
\eea
which shows the same dependence of the size in the long axis as in a uniform distribution of a needle-like geometry \cite{Rehler1971}. Again by using the total volume $V$$=$$\pi^{3/2}R_\perp^2R_L$ of Gaussian density distributions and the averaged density $\rho$$=$$N/V$, we derive the cooperative linewidth as
\bea
\tilde\gamma_{31}/\gamma_{31}=\frac{3\sqrt{\pi}}{8}\frac{N}{|\k_p|R_L}=\frac{\pi}{8}D_c\frac{|\k_p|R_\perp^2}{2R_L},
\eea
where $D_c$ $=$ $2\rho\sigma R_L$ is the effective optical depth. We note that the relevant parameters $\rho$, optical depth $D_c$, a total number of atom $N$, $R_\perp$, and $R_L$ deterministically characterize the cooperative phenomena and enables a direct connection with EIT transmission measurements. These parameters are not independent from each other, since $\rho=N/V$ with $V=\pi^{3/2}R_\perp^2R_L$ and optical depth $D_c=2\rho\sigma R_L$. 

\begin{figure}[t]
\centering
\includegraphics[width=8.5cm,height=4.5cm]{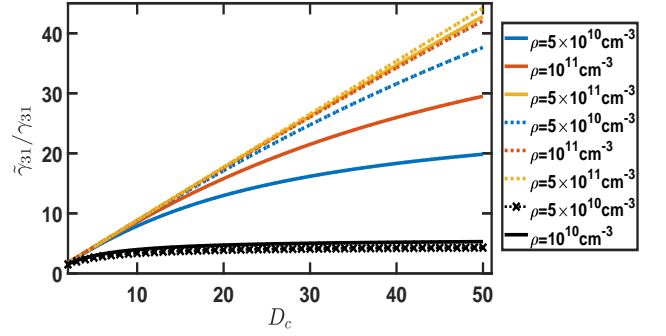}
\caption{Linewidth broadening from RDDI. We plot the enhanced decay rates $\tilde \gamma_{31}$ normalized by an intrinsic decay constant $\gamma_{31}$ for various atomic densities with $R_\perp=50$ (dot), $25$ (solid), and $10~\mu$m ($\times$). At a relatively low atomic density, $\tilde\gamma_{31}$ gradually saturates as $D_c$ increases, while at a large density, it approaches a linear dependence on $D_c$. For a larger $R_\perp$, linewidth broadens more significantly.}\label{fig1}
\end{figure}

In Fig. \ref{fig1}, we demonstrate $\tilde\gamma_{31}$ of Eq. (\ref{general}) for three different $R_\perp$. For a fixed $\rho$, $\tilde\gamma_{31}$ saturates as $D_c$ increases for a small $\rho$ while approaches a linear dependence for a larger $\rho$. For a smaller $R_\perp$, $\tilde\gamma_{31}$ appears less significant due to small $N$ or $V$ involved for the cooperative linewidth at the same $D_c$. The interplay between $R_\perp$ and $\rho$ can be hardly distinguished in $\tilde\gamma_{31}$, which we show in the upper bounds of Fig. \ref{fig1}, where $\rho=1-5\times 10^{11}$cm$^{-3}$ with a larger $R_\perp$ almost overlaps with the case of $\rho=5\times 10^{11}$cm$^{-3}$ with a halved $R_\perp$. We also show almost identical lower bounds, as an example, for the cases of low $\rho$ and smaller $R_\perp$ in Fig. \ref{fig1}. This reflects the difficulty in determining precise atomic configurations of $N$, $\rho$, or $D_c$ straightforwardly from the cooperative linewidth $\tilde\gamma_{31}$, and therefore other complimentary and independent measurements are necessary to ensure these crucial parameters relevant to the linewidth of the probe field. As for extreme needle-like geometry, we find a suppressed $\tilde\gamma_{31}$ due to the factor of $|\k_p|R_\perp^2/R_L$. 

We note that if $\rho$ and $N$ are fixed (so is $V$), increasing $R_\perp$ would degrease the line broadening as there would be less atoms contributing to the enhanced decay rate along the propagation direction. This can be seen if we take $\rho=5\times 10^{10}$ cm$^{-3}$ as an example in Fig. \ref{fig1}. As $R_\perp$ increases from $25$ to $50$ $\mu$m, $D_c= 50$ decreases to $12.5$ for the same $\rho$, $N$, and $V$. The corresponding $\tilde{\gamma}_{31}$ in Fig. \ref{fig1} reduces from $20$ to $12$. 

For cooperative frequency shift \cite{Scully2009,Jen2015}, it is related to cooperative spontaneous decay rate, fulfilling Kramers–Kronig relations in electric susceptibility. We introduce an infrared cutoff of wave vector $k_m$ $=$ $2\pi/\sqrt[3]{V}$ in calculating the cooperative frequency shift \cite{Jen2015}, and we have
\bea
\tilde\delta_p=\delta_p-\frac{2}{\sqrt{\pi}}(\tilde\gamma_{31}-\gamma_{31})\frac{\lambda}{\sqrt[3]{R_\perp^2 R_L}},
\eea
with the probe field transition wavelength $\lambda$. For $R_\perp$ $=$ $50$ $\mu$m and $R_L$ $=$ $1$ mm, we estimate the cooperative frequency shift $\sim$ $\tilde\gamma_{31}/154.2$ for a moderate atomic density we consider in Fig. \ref{fig1}, which has less effect than the cooperative linewidth does in EIT spectroscopy.

\section{Multiple scattering of RDDI in transmission}

\begin{figure}[t]
\centering
\includegraphics[width=8.5cm,height=4.5cm]{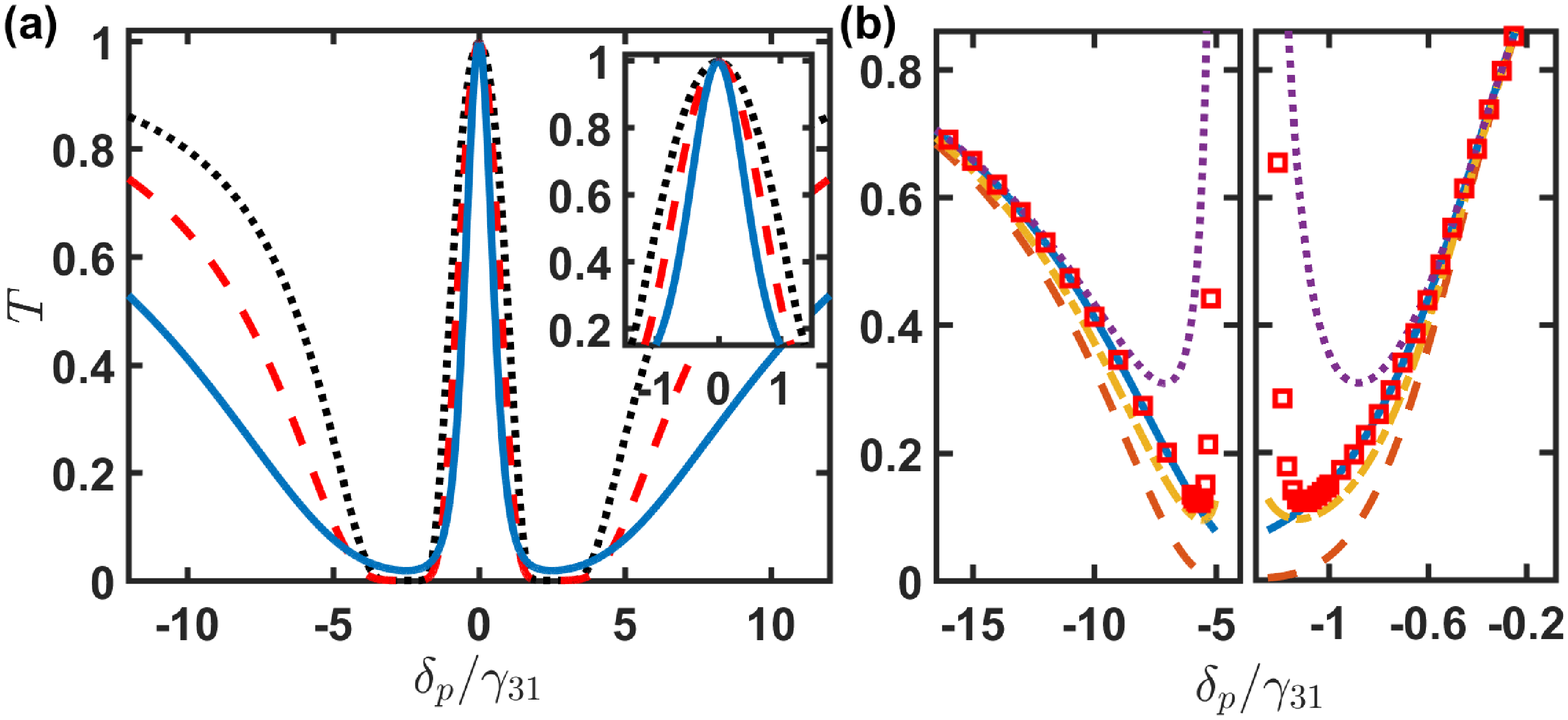}
\caption{EIT transmission and convergence of multiple scattering of RDDI. (a) The transmission $T$ near two-photon resonance (inset) has shorter probe spreads from $\tilde\gamma_{31}/\gamma_{31}$ $=$ $1$ (dot), $2$ (dash) to $5$ (solid), with $D_c=20$, $\delta_c$ $=$ $0$, $\Omega_c=5\gamma_{31}$, and $\gamma_{21}/\gamma_{31}=0.001$. The absorption peaks become broader at $|\delta_p|\gtrsim\Omega_c$ as $\tilde\gamma_{31}$ increases. (b) For the case of $\tilde\gamma_{31}/\gamma_{31}=5$ (solid), the multiple scattering of RDDI in $T_M$ shows convergence inside the transparency window and at off-resonance in the right and left panels respectively, with scattering orders $M=1$ (dash), $2$ (dash-dot), $3$ (dot), and $20$ ($\square$).}\label{fig2}
\end{figure}

Next we directly apply the results of $\tilde\gamma_{31}$ in the previous section from a general cigar-shaped geometry to the transmission property we obtain in Sec. II. Specifically, we compare the transmission property $T$ of Eq. (\ref{transmission}) under all-order scattering of RDDI to the one under a finite $M$th-order multiple scattering effect in Eq. (\ref{transmission_order}). With this comparison shown in Fig. \ref{fig2}, we are able to present the convergence of multiple scattering of RDDI and uncover the parameter regions in EIT spectroscopy where our perturbative treatment is most valid. 

We first plot $T$ of Eq. (\ref{transmission}) in Fig. \ref{fig2}(a). As $\tilde\gamma_{31}$ increases, the transmission window allows for shorter probe frequency spreads, and thus puts more restrictions in light propagation and reduces storage efficiency. Larger cooperative linewidth also broadens the absorption peaks. We have absorbed the cooperative frequency shifts into $\delta_p$, which nonetheless can also be observable in conventional EIT experiments. Next in Fig. \ref{fig2}(b), we further compare the transmission with finite orders of multiple scattering of RDDI in Eq. (\ref{transmission_order}). A clear convergence emerges toward higher order of perturbations, both near transparency and in the off-resonance regimes. This demonstrates that our result of transmission $T$ from Eq. (\ref{transmission}) essentially involves all-order scattering of RDDI, which, therefore, enables a direct comparison with present experiments. Meanwhile, close to the transmission valley (absorption peaks), a divergence shows up and the perturbation fails in this region of EIT spectrum.

\begin{figure}[t]
\centering
\includegraphics[width=8.5cm,height=4.5cm]{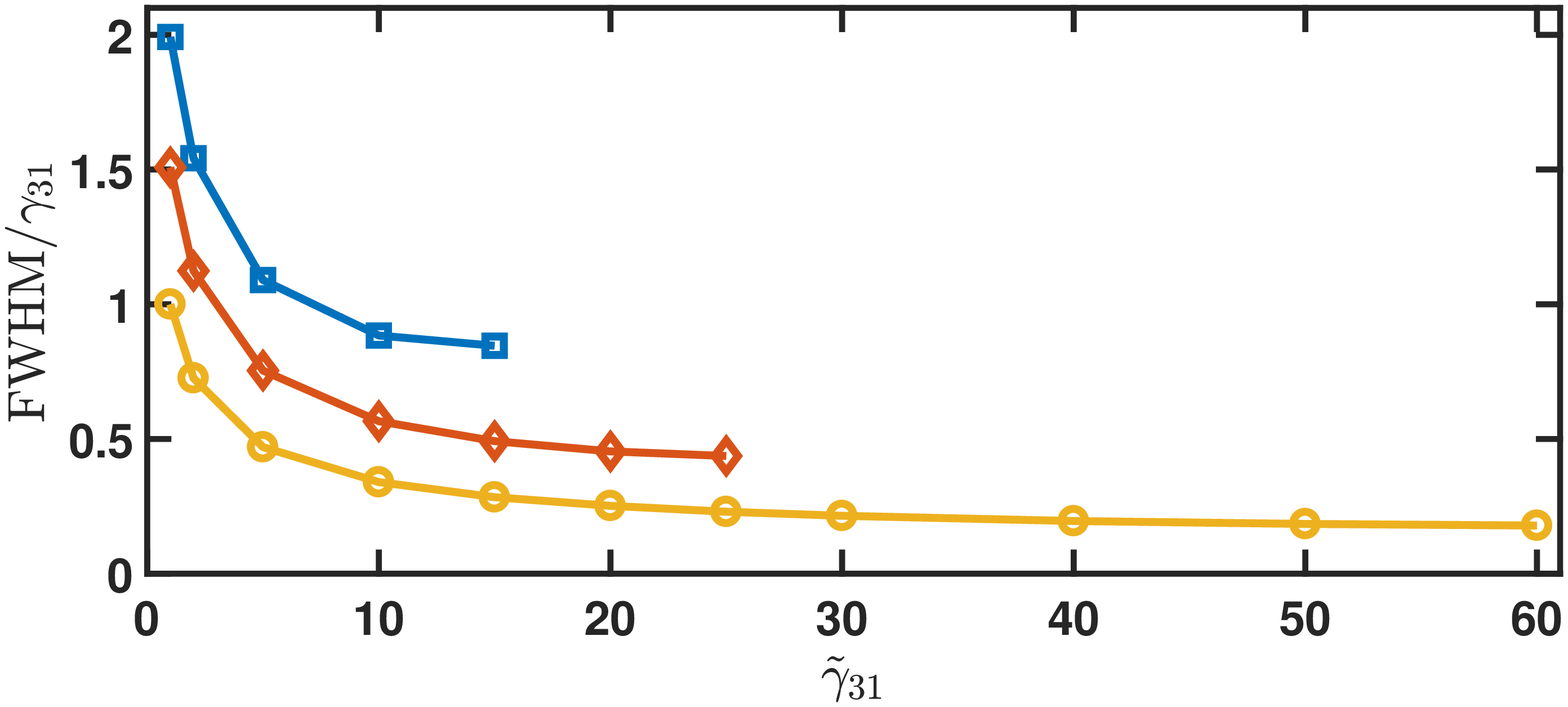}
\caption{Full-width-half-maximum (FWHM) of EIT transmission window. We plot the FWHM of the transparency window for $D_c=20$ ($\square$), $40$ ($\diamond$), and $100$ ($\circ$), for various cooperative linewidth broadening $\tilde\gamma_{31}$. Other parameters of EIT setup are the same as Fig. \ref{fig2}.}\label{fig3}
\end{figure}

Finally, in Fig. \ref{fig3} we plot the FWHM of transparency window as a dependence of cooperative linewidth $\tilde\gamma_{31}$ for various optical depths $D_c$. As the optical depth increases, this bandwidth decreases as expected since in the limit of large $\Omega_c$ but $\Omega_c\ll \sqrt{D_c}\Gamma$, it becomes \cite{Lukin1997, Wei2020}  
\bee
{\rm FWHM}=\sqrt{\frac{\log(2)}{2}}\frac{\Omega_c^2}{\sqrt{D_c\Gamma\tilde{\gamma}_{31}}},
\eee
which can be fitted by a Gaussian function of the detuning $\delta_p$. In contrast, for a finite $\Omega_c$, as the cooperative linewidth increases, the bandwidth decreases but saturates for large $D_c$. In Fig. \ref{fig3}, we stop the bandwidth calculations at a certain $D_c$ where Eq. (\ref{transmission}) no longer genuinely determines the FWHM of the transparency regions. Furthermore, we find a scaling of $\tilde\gamma_{31}^{-a}$ with $a=0.34$, $0.41$, and $0.45$, respectively for $D_c=20$, $40$, and $100$. This presents a significant reduction of the FWHM of EIT transmission window in a more optically thick atomic ensemble.

In contrast for Rydberg EIT experiment \cite{Pritchard2010} and theory \cite{Petrosyan2011}, the transparency is suppressed due to Rydberg dipole-dipole interactions. This nonlinear interaction results in an effective photon-photon interaction, which can be applied in realizing photonic quantum gates and forming many-body photonic bound states. On the other hand, in EIT with RDDI here, the transparency does not change significantly since no photon-photon interaction is present. What is significantly modified is the transparency bandwidth which is narrowed due to the finite cooperative linewidth. Moreover, the off-resonance absorption peaks are broadened, which can, along with the information of decreasing transparency bandwidth, further characterize the cooperative effect of RDDI in EIT.


\section{Beyond local field approximation} 

Here in the last section of our main results, we release the local field approximation, where multiple nonlocal scattering events of RDDI can be envisaged in EIT. In practical experiments for a large optical depth, $K_{\alpha\beta}$ can be order of $\gamma_{31}$, and such that a perturbative treatment of $K_{\alpha\beta}$ is only valid when RDDI is weak in a cold gas with a low density or near the transparency regions in EIT scheme. We go beyond this approximation and proceed from Eqs. (\ref{MB}) and (\ref{sigma13}). We reevaluate the steady-state solutions still under the condition of a weak $\Omega_p$, and so we keep only the first order of $\Omega_p$ in the Maxwell-Bloch equations of Eq. (\ref{MB}). 

For the first extension, we include the transverse part of field dynamics, which we will see later that it would be necessary when RDDI becomes significant. Equation (\ref{MB}) becomes
\bea
\left(-\frac{i}{2k_p}\nabla^2_\perp+\frac{\partial}{\partial z}\right)\Omega_p(\r)=&&\frac{iD\Gamma}{2L}\tilde{\sigma}_{13}(\r),\label{MB2}
\eea
where $\nabla^2_\perp$ denotes the vector Laplacian in transverse directions of $\hat x$ and $\hat y$. The above presents a three-dimensional propagation equation (note that we use $\r$ for the field and the dipole operator) with transverse dynamics, which takes care of refraction if there is strong RDDI. 

The second extension goes to Eq. (\ref{sigma13}), where we release the cross-section average and obtain
\bea
0=&&\left[i\delta _{p}-\gamma_{31}-\frac{i|\Omega_c|^2}{4(\delta_2+i\gamma_{21})}\right]\tilde{\sigma}_{13}^\alpha+i\frac{\Omega _{p}(\r_\alpha)}{2} \nonumber\\
&&-\sum_{\beta\neq\alpha}^NK_{\alpha\beta}\tilde\sigma_{13}^\beta,
\eea
where we have substituted Eq. (\ref{sigma12_steady}). From the above, we further obtain, in discrete forms in space, 
\bea
i\hat M \vec\sigma_{13}=&&-\frac{\vec\Omega_p}{2},\\
\vec\sigma_{13}=&&\frac{i}{2}\hat M^{-1}\vec\Omega_p,\label{sigma13_2}
\eea
where $\vec \sigma_{13}$ $=$ $[\sigma_{13}(r_1),$ $ \sigma_{13}(r_2)$ , ..., $\sigma_{13}(r_N) ]$, $\vec\Omega_p$ $=$ $[\Omega_p(r_1)$, $\Omega_p(r_2)$, ..., $\Omega_p(r_N)]$, and 
\bea
M=\begin{bmatrix}
    -A  & K_{1,2} & K_{1,3} & \dots & K_{1,N}\\
    K_{2,1} & -A & K_{2,3} & \dots & K_{2,N}\\
		K_{3,1} & K_{3,2}  & -A  & \dots  & \vdots  \\
    \vdots 	& \vdots  & \vdots & \ddots & K_{N-1,N}  \\
		K_{N,1} & \dots & \dots & K_{N, N-1} & -A
\end{bmatrix},\label{M}
\eea
with $A= i\delta _{p}-\gamma_{31}-i|\Omega_c|^2/[4(\delta_2+i\gamma_{21})]$ which is also defined in Sec. II. From $\hat M$, we can see that the dipole operators are induced by or emerged from coupling to nonlocal fields. 

With Eqs. (\ref{MB2}) and (\ref{sigma13_2}) together, they describe propagation dynamics of a probe field with the effect of RDDI. By combining Eqs. (\ref{MB2}) and (\ref{sigma13_2}) in continuous limit, we obtain 
\bea
\left(-\frac{i}{2k_p}\nabla^2_\perp+\frac{\partial}{\partial z}\right)\Omega_p(\r)=&&-\frac{D\Gamma}{4L}\int\hat M^{-1}(\r-\r')\nonumber\\
&&\times \Omega_p(\r')\rho(\r')d\r',\label{nonlocal}
\eea
where $\rho(\r)$ is the atomic density. The above results show that probe field dynamics can be influenced by RDDI in both longitudinal and transverse directions. It reduces to conventional EIT in single-particle picture when $K_{\alpha,\beta}\rightarrow 0$, leading to a local field propagation equation dominated simply by an electric susceptibility $\chi\propto A^{-1}$. 

From Eq. (\ref{nonlocal}), we note that a local field approximation we apply in perturbative treatments is valid only when the field is near the transparency or non-resonant regimes. This means that when the probe field $\Omega_p(\r)$ at arbitrary positions $\r$ is not significantly attenuated and almost sustains its incident waveform at $\r=0$, that is $\Omega_p(\r)\approx\Omega_p(\r=0)$. Under this condition, the effect of RDDI on the effective decay rate of probe transition is most significant, since the integral in Eq. (\ref{nonlocal}) goes through all the atoms. On the other hand, when the probe field is strongly attenuated, according to Eq. (\ref{nonlocal}), no significant RDDI manifests in modifying the probe field dynamics. Therefore, to determine the full EIT spectrum and associated linewidth of EIT window, a single quantity of effective decay rate is not sufficient owing to this asymmetry of regimes, especially for an atomic gas with a high optical density. 

We can formally solve Eq. (\ref{nonlocal}) by Laplace transforms since it has convolution forms between probe field and $\hat M^{-1}$. We then obtain (first neglecting the refraction terms and assuming a homogeneous gas $\rho=N/V$),
\bea
s\bar\Omega_p(s)-\Omega_p(z=0)=-\frac{D\Gamma}{4L}(N/L)\bar M^{-1}(s)\bar\Omega_p(s),
\eea 
where $F(s)\equiv\int_0^\infty f(r)e^{-sr}dr$ and $s$ represents a complex number as for complex momentum spaces in Laplace transforms of real spatial spaces $r$ and $r'$. This further gives
\bea
\bar\Omega_p(s)=\frac{\Omega_p(z=0)}{s+\frac{D\Gamma}{4L}(N/L)\bar M^{-1}(s)},
\eea
which can be inverse transformed in the complex plane back to $\Omega_p(z=L)$ by
\bea
\Omega_p(z=L)=\frac{1}{2\pi i}\oint_{r-iR}^{r+iR}\bar\Omega_p(s) e^{sL}ds.
\eea

To further solve the above $\Omega_p(z=L)$, we assume again a perturbative RDDI, such that 
\bea
M^{-1}=(-A\hat I+\hat K)^{-1}\approx \frac{-1}{A}\left(\hat I+\frac{\hat K}{A}\right),
\eea
where $\hat K$ represents the off-diagonal elements in Eq. (\ref{M}). Now a weak RDDI can be approximated by a far-field expression in $K(r-r')\rightarrow (3\Gamma/2)(-ie^{i\xi}e^{-i\k_p\cdot(\r-\r')})/\xi$. We finally obtain
\bea
&&\frac{\Omega_p(z=L)}{\Omega_p(z=0)}\nonumber\\
&&=\frac{1}{2\pi i}\oint \frac{e^{sL}ds}{s-\frac{D\Gamma}{4AL}-\frac{D\Gamma}{4AL}(\frac{N}{k_pL})(\frac{3\Gamma/2}{A})e^{as}\bar\Gamma(0,as)},\label{sT}
\eea
where $a$ is introduced in Laplace transform of $1/(|r-r'|+a)$ to remove the divergence when $r\rightarrow r'$ and is in unit of space as a complex length scale to renormalize the RRDI at divergence, and $\bar\Gamma$ is an incomplete Gamma function. The above is a square root of transmission, and the second term in the denominator of Eq. (\ref{sT}) is exactly the residue of the complex integral, leading to the EIT transmission ($T=|\Omega_p(z=L)/\Omega_p(z=0)|^2$) without RDDI,
\bea
\log\left(\frac{\Omega_p(z=L)}{\Omega_p(z=0)}\right)=\frac{D\Gamma}{4A}.\nonumber
\eea 
Interestingly, the third term in the denominator of Eq. (\ref{sT}) shows the RDDI effect leading to multiple residues in the complex plane, which indicates the failure of expressing EIT spectrum in terms of single effective decay rate. We note that these multiple scales in decay constants show up even for a weak RDDI we consider here. In general, the multiple contributions of collective resonances and decay constants are most evident in low dimensional and dense atomic arrays, and the field propagating through constituent particles can not be genuinely described by a typical Beer–Lambert law \cite{Chomaz2012} or Maxwell-Bloch equation in a macroscopic media. 

\section{Discussion and conclusion}

Generally speaking, the RDDIs are universal in all light-matter interacting systems, which are most prominent in a dense media. They are responsible for many intriguing phenomena of superradiance and subradiance, and can be exploited to significantly enhance the performance of quantum storage efficiency in EIT under some tailored collective states \cite{Garcia2017}. Both these cooperative spontaneous emissions emerge from strong RDDIs, where multiple scattering of light exchanges between these quantum emitters dominate the dissipation process of the excited atoms. 

This universal effect of RDDI should be observable as well in the EIT media, but a systematic and evident study on the collective effect from RDDI has not been reported yet or fully accounted for in most of the theoretical investigations. We note that there are two central assumptions in our theory: weak field excitation and local field approximation. The first assumption gives us the lowest level of hierarchy in the coupled equations, which also associates with the first-order cumulant expansion we apply in this work and ignorance of higher-order atom-atom correlations that would be relevant in dense media. The second assumption allows us to obtain the analytical forms of the RDDI effect on EIT properties, without which resolving the RDDI effect would otherwise require full eigen-spectrum of the system. We think it is the local field approximation that leads to a dramatic RDDI effect in our theoretical treatments, where the contributions of RDDI throughout the whole atomic ensemble are included at the local field. Therefore, our predicted RDDI effect here could be overestimated compared to nowadays EIT experiments. Furthermore, we elaborate that a further advanced but computationally exhaustive quantum Langevin approach may provide a way to get around our assumptions here and take full account of light propagation, system dynamics, and atom-atom correlations in large $N$ limit.

For highly efficient EIT-based quantum memory applications \cite{Cho2016, Gris2018, Hsiao2018}, the atomic system is often prepared to be elongated along the propagation direction, leading to a high optical depth for the probe light. In these high optical density experiments, there is no clear observation of the RDDI effect on EIT spectroscopy, except in \cite{Hsiao2018} a mild dependence of an optical depth on the cooperative line width is conjectured in determining the experimental parameters by fitting EIT spectra and slow light traces. The fitted decay rate for the probe field transition has been shown to be enhanced two to three times as the optical density increases, which suggests the cooperative effect of RDDI in EIT. As a comparison to our predictions in Fig. \ref{fig2}, we have $\tilde \gamma_{31}/\gamma_{31}\sim 27$ and $4.5$ for $R_\perp=25$ and $10~\mu$m, respectively, at $\rho=5\times 10^{10}$cm$^{-3}$ with $D_c=400$. This shows the overestimation of our model and quite a deviated range of predicted collective decay rates owing to different geometries of the atoms.  

To clearly identify the RDDI effect with genuine comparisons, a further systematic investigation is needed to (i) accurately determine the number of atoms and the associated interacting volume along with its optical density and (ii) acquire weak enough or a single photon source to serve as a probe field excitation. This way, our theory here can provide a direct comparison to the EIT experiments to potentially unravel the RDDI effect on the line narrowing of the transmission window. It seems that a mild optical depth and standard atomic density in our theoretical consideration in Figs. \ref{fig2} and \ref{fig3} already suggests a clear signature of RDDI. We attribute this clear signature to our perturbative treatments using local field approximation. A complexity can arise when this approximation is released as we have demonstrated in Sec. V, where multiple collective resonances and decay rates may participate in determining EIT transmission. We also note that the role of atom-atom correlations from RDDI can be crucial in light scattering of dense atoms \cite{Jennewein2016, Robicheaux2020}, and they should as well be non-negligible in a large $N$ and high optical depth system we consider here. To include this effect of light-induced atom-atom correlations which are not accounted for in our theoretical treatment, either the next-order cumulants can be included or quantum Langevin equation in positive-$P$ phase space \cite{Drummond1987, Gardiner2000, Jen2012} can be conducted to go beyond our perturbative treatments here.   

As for Langevin equations on the spontaneous emissions \cite{QO:Scully}, quantum fluctuations or quantum noises play the role in initiating the dissipations of the excited atoms. In EIT, since the probe field intensity is under the normal order of field operators, it is commonly believed that the quantum fluctuations do not play a role in the probe field transmission \cite{Lukin2003, Fleischhauer2005}. We note here that one of the crucial assumption in previous theoretical methods is the weak probe field approximation. Under this perturbation of the probe fields, up to single atomic excitation is allowed, and the collective spin excitation can form and lead to the popular picture of dark-state polariton. This picture sustains the bosonic commutation relations of the quasi-particles in large $N$ limit and can explain some essential features of quantum storage and retrieval of single photon propagation in EIT media. We emphasize that this assumption can be easily broken when the light-matter interacting system goes beyond the single excitation limit when multiple excitations are present. When we have $M$ atomic excitations in the atomic ensemble with $M$ $\lesssim$ $N$, naturally we face an exponential growth of the number of the states in the dynamical Hilbert space in the order of $C^N_M$ $\sim$ $\mathcal{O}(N^M)$ if $M$ $\ll$ $N$. This indicates a dilemma and challenge we would encounter in any quantum systems when a full capacity of light-matter interactions is considered in describing their system dynamics. 

A relevant platform which deviates from the conventional EIT setup and has distinct EIT properties is the Rydberg EIT scheme \cite{Saffman2010, Pritchard2010, Dudin2012, Peyronel2012, Hsiao2020, Kim2021}. In this platform which utilizes the high-lying Rydberg excited states, significant nonlinear effect of nonlocal dipole-dipole interactions emerge and reduce the performance of the probe field transparency. This interaction leads to the dipole blockade that forbids Rydberg excitations within the blockade radius, along with a significant energy shift away from the conventional transparency condition of two-photon resonance. Here instead we investigate the RDDI effect in the conventional EIT setup, where the dipole-dipole interactions are induced from the light exchanges between the atomic levels of the ground and the first excited states, even though RDDIs by nature have the feature of nonlocal interaction, similar to the Rydberg EIT scheme. By contrast, the dipole-dipole interactions in Rydberg EIT media directly result from the Rydberg excited levels. The theory of Rydberg EIT takes the strategy of single-particle picture with a mean-field average of the Rydberg interactions, which results in an effective nonlocal and nonlinear interaction in EIT spectrum. Two recent studies also present a demand of a more complete understanding of Rydberg EIT system \cite{Bienias2020, Tebben2021}, where radiation trapping of scattered photons \cite{Bienias2020} or emerging nontrivial photon correlations \cite{Tebben2021} might be crucial to fill the gap between the mean-field models and experiments.  

In conclusion, we have theoretically investigated the role of RDDI in EIT properties. We have predicted the effective cooperative linewidth and frequency shift in EIT transmission from the multiple scattering of RDDI, which provides a direct comparison with experiment on the cooperative phenomena in EIT. The allowed transparent probe spread in the EIT transmission window is reduced due to a finite cooperative linewidth, which makes light storage less efficient in EIT-based quantum memory application. The phenomenon of RDDI in EIT should be observable in conventional EIT experiments in atoms with a moderate atomic density and optical depth. Finally, we note that our work here is just a starting point of the study of RDDI in EIT. The discrepancy between the theory and experiment is noted and certainly deserves further exploration.

\section*{Acknowledgments}

This work is supported by the Ministry of Science and Technology (MOST), Taiwan, under the Grant No. MOST-109-2112-M-001-035-MY3. We are also grateful for support from Thematic Group 1.2 and 3.2 of National Center for Theoretical Sciences in Taiwan.
\appendix
\section{Maxwell-Bloch equations for EIT with RDDI} 

In this section, we derive the equations of motion for EIT with cooperative effects in the probe transition. We first define the slow-varying coherence operators as
\bea
\tilde\sigma_{13}(z,t)=&&\frac{1}{N_z}\sum_{\beta=1}^{N_z}\hat\sigma_{13}^\beta(z,t) e^{i\omega_pt-i\k_p\cdot\r_\beta},\\
\tilde\sigma_{23}(z,t)=&&\frac{1}{N_z}\sum_{\beta=1}^{N_z}\hat\sigma_{23}^\beta(z,t) e^{i\omega_ct-i\k_c\cdot\r_\beta},\\
\tilde\sigma_{12}(z,t)=&&\frac{1}{N_z}\sum_{\beta=1}^{N_z}\hat\sigma_{12}^\beta(z,t) e^{i\Delta\omega t-i\Delta \k\cdot\r_\beta},
\eea
where $\Delta \k$ $=$ $\k_p$ $-$ $\k_c$, and $\Delta\omega$ $=$ $c|\Delta \k|$.\ Same cross section average is applied for the population operators,
\bea
&&\tilde\sigma_{11}=\frac{1}{N_z}\sum_{\beta=1}^{N_z}\hat\sigma_{11}^\beta(z,t),~
\tilde\sigma_{22}=\frac{1}{N_z}\sum_{\beta=1}^{N_z}\hat\sigma_{22}^\beta(z,t),\nonumber\\
&& \tilde\sigma_{33}=\frac{1}{N_z}\sum_{\beta=1}^{N_z}\hat\sigma_{33}^\beta(z,t).\nonumber
\eea

These cross-grained averages of the slow-varying variables are typical treatments in solving the propagating quantized electric fields through an atomic medium. Later for predictions of EIT properties, we shall not encounter the effect from cross-grained details, which is true since we take $N$ and $N_z$$\rightarrow$$\infty$. In the following, we will consider only one-body atomic operators in the dynamically light-matter coupled equations, which truncates the hierarchy of the equations and equivalently neglects small higher order multi-atom correlations.     

In the co-propagating frame $\tau$ $=$ $t$ $-$ $z/c$, we obtain the Maxwell-Bloch equations as
\begin{widetext}
\bea
\frac{d }{d \tau}\tilde{\sigma}_{23} =&&(i\delta _{c}-\gamma_{21}-\gamma_{32})\tilde{\sigma}_{23}+i\frac{\Omega _{c}}{2}(\tilde{\sigma}_{22}-\tilde{\sigma}_{33})+i\frac{\Omega_p}{2}\tilde{\sigma}_{12}^{\dag},\label{sigma23}\\
\frac{d}{d\tau}\tilde{\sigma}_{13} =&&(i\delta _{p}-\gamma_{31})\tilde{\sigma}_{13}+i\frac{\Omega _{c}}{2}\tilde{\sigma}_{12}+i\frac{\Omega_p}{2}(\tilde{\sigma}_{11}-\tilde{\sigma}_{33})-\frac{1}{N_z}\sum_{\alpha=1}^{N_z}\sum_{\beta\neq\alpha}^NK_{\alpha\beta}(\tilde\sigma_{11}^\alpha-\tilde\sigma_{33}^\alpha)\tilde\sigma_{13}^\beta,\label{sigma13_app}\\
\frac{d}{d \tau }\tilde{\sigma}_{12} =&&(i\delta_2-\gamma_{21})\tilde{\sigma}_{12}-i\frac{\Omega_p}{2}\tilde{\sigma}_{23}^{\dag}+i\frac{\Omega_{c}^{\ast}}{2}\tilde{\sigma}_{13},\\
\frac{d}{d\tau }\tilde{\sigma}_{11} =&&2\gamma _{31}\tilde{\sigma}_{33}+2\gamma _{21}\tilde{\sigma}_{22}-i\frac{\Omega_p}{2} \tilde{\sigma}_{13}^{\dag}+i\frac{\Omega_p^*}{2}\tilde{\sigma}_{13}+\frac{1}{N_z}\sum_{\alpha=1}^{N_z}\sum_{\beta\neq\alpha}^N\left[K_{\alpha\beta}(\tilde\sigma_{31}^\alpha\tilde\sigma_{13}^\beta)+K_{\alpha\beta}^*(\tilde\sigma_{31}^\beta\tilde\sigma_{13}^\alpha)\right],\\
\frac{d}{d\tau }\tilde{\sigma}_{33} =&&-\gamma _{3}\tilde{\sigma}_{33}+i\frac{\Omega_p}{2} \tilde{\sigma}_{13}^{\dag}-i\frac{\Omega_p^*}{2}\tilde{\sigma}_{13}+i\frac{\Omega_c}{2}\tilde{\sigma}_{23}^{\dag }-i\frac{\Omega_c^*}{2}\tilde{\sigma}_{23}-\frac{1}{N_z}\sum_{\alpha=1}^{N_z}\sum_{\beta\neq\alpha}^N\left[K_{\alpha\beta}(\tilde\sigma_{31}^\alpha\tilde\sigma_{13}^\beta)+K_{\alpha\beta}^*(\tilde\sigma_{31}^\beta\tilde\sigma_{13}^\alpha)\right],\\
\frac{d}{d z}\Omega_p=&&\frac{iD\Gamma}{2L}\tilde{\sigma}_{13},\label{MB}
\eea
\end{widetext}
where $K_{\alpha\beta}$ $\equiv$ $(F_{\alpha\beta}$ $+$ $i2G_{\alpha\beta})e^{-i\k_p\cdot\r_{\alpha\beta}}/2$ \cite{Lehmberg1970}. The slow-varying atomic operators, not the cross-grained ones, are denoted as $\tilde{\sigma}_{\mu\nu}^\beta$, and for example, $\tilde{\sigma}_{13}^\beta$ $=$ $\hat\sigma_{13}^\beta e^{i\omega_pt-i\k_p\cdot\r_\beta}$. The optical depth is $D$ $\equiv$ $\rho\sigma L$ and $\gamma_3$ $=$ $2\gamma_{31}$ $+$ $2\gamma_{32}$. The two-photon detuning is $\delta_2$ $=$ $\delta_p$ $-$ $\delta_c$. The RDDI in the probe transition modifies the transition coherence $\tilde\sigma_{13}$ and redistributes the populations of $\tilde\sigma_{11}$ and $\tilde\sigma_{33}$ through multi-atom correlations.

In this $\Lambda$-type atomic system, we assume all $N$ atoms are initially prepared in $|1\rangle$.\ In the limit of weak probe field, which corresponds to a linear dependence of $\Omega_p$ in the coupled equations, we have $\tilde\sigma_{11}$ $=$ $1$, $\tilde\sigma_{22}$ $=$ $\tilde\sigma_{33}$ $=$ $0$, and $\tilde\sigma_{23}$ $=$ $0$.\ The relevant equations for atomic coherences (with implicit spatial dependence of $z$) are
\bea
\frac{d}{d \tau }\tilde{\sigma}_{12} \approx&& (i\delta_2-\gamma_{21})\tilde{\sigma}_{12}+i\frac{\Omega_{c}^{\ast}}{2}\tilde{\sigma}_{13},\label{sigma12}\\
\frac{d}{d\tau}\tilde{\sigma}_{13} \approx&& (i\delta _{p}-\gamma_{31})\tilde{\sigma}_{13}+i\frac{\Omega _{c}}{2}\tilde{\sigma}_{12}+i\frac{\Omega_p}{2}\nonumber\\
&& -\frac{1}{N_z}\sum_{\alpha=1}^{N_z}\sum_{\beta\neq\alpha}^NK_{\alpha\beta}\tilde\sigma_{13}^\beta.\label{sigma13}
\eea 

\section{Steady state solutions}

Here we proceed to solve the steady-state solutions of the coupled equations in Appendix A. 2 and present how the effect of RDDI modifies the EIT spectrum. The steady state solution of the ground state coherence from Eq. (\ref{sigma12}) gives
\begin{equation}
\tilde \sigma_{12}=\frac{-\Omega_c^*/2}{\delta_2+i\gamma_{21}}\tilde\sigma_{13}.\label{sigma12_steady}
\end{equation}
We then substitute $\tilde \sigma_{12}$ from the above in Eq. (\ref{sigma13}), and in the zeroth order of $K_{\alpha\beta}$, for a real constant $\Omega_c$, we derive the coherence of the probe transition (now retrieving the spatial dependence for clarity)
\bee
\tilde\sigma_{13}^{(0)}(z)=\frac{-i\Omega_p(z)/2}{i\delta_p-\gamma_{31}-\frac{i\Omega_c^2/4}{\delta_2+i\gamma_{21}}}\equiv\frac{-i\Omega_p(z)/2}{A},\label{zero}
\eee
which is proportional to the electric field susceptibility in conventional EIT theory without cooperative effects. We define $A$ above for later concise representation of the cooperative effect on the linewidth of EIT spectrum. 

Next for the first-order perturbation of $K_{\alpha\beta}$, we apply the zeroth order results of Eq. (\ref{zero}) to Eq. (\ref{sigma13}) and obtain
\bee
\tilde\sigma_{13}^{(1)}(z)=\frac{-i\Omega_p(z)/2}{A}+\frac{1}{AN_z}\sum_{\alpha=1}^{N_z}\sum_{\beta\neq\alpha}^NK_{\alpha\beta}\tilde\sigma_{13}^{\beta,(0)}(\r_\beta),\label{DD}
\eee
where a coupling between the atomic coherences at $z$ and other positions $\r_\beta$ $\neq$ $\r_\alpha$ appears due to RDDI. Equation (\ref{DD}) essentially describes nonlocal interactions of the fields. We proceed to apply the local field approximation to Eq. (\ref{DD}) and obtain
\bee
\tilde\sigma_{13}^{(1)}(z)=\frac{-i\Omega_p(z)/2}{A}+\frac{f_{C}}{A}\frac{-i\Omega_p(z)/2}{A},\label{cooperative}
\eee
where $f_{C}$ $=$ $N_z^{-1}\sum_{\alpha=1}^{N_z}\sum_{\beta\neq\alpha}^{N}K_{\alpha\beta}$ denotes the cooperative dipole-dipole interactions between the cross-grained region and all the other atoms in the ensemble. We calculate the leading order of $f_{C}$ in the main paper, which is independent of the cross-grained region and is cooperatively enhanced due to the involvement of all $N$ atoms of the medium.

From light propagation of Eq. (\ref{MB}) in the Maxwell-Bloch equations, we replace the above $\tilde{\sigma}_{13}$ with $\tilde\sigma_{13}^{(1)}(z)$ of Eq. (\ref{cooperative}), and $\Omega_p(L)$ becomes 
\bee
\log\left(\frac{\Omega_p(L)}{\Omega_p(0)}\right)=\frac{D\Gamma}{4}\left(\frac{1}{A}+\frac{f_{C}}{A^2}\right)
\approx\frac{D\Gamma}{4}\frac{1}{A-f_{C}},\label{modified}
\eee
where the approximate form in the above is valid when $|f_C|\ll |A|$, and it suggests that there is an effective $\tilde\delta_p^{(1)}$ and $\tilde\gamma_{31}^{(1)}$ in $A$, which are
\bee
\tilde\delta_p^{(1)}=\delta_p - \textrm{Im}[f_{C}],~
\tilde\gamma_{31}^{(1)}=\gamma_{31}+\textrm{Re}[f_{C}].\label{LW}
\eee
These represent a cooperative frequency shift in $\delta_p$ and linewidth broadening (larger than $\gamma_{31}$) respectively, and their superscripts denote the single scattering event of RDDI.

To account for the multiple scattering effect in EIT theory, similar to the treatment of coherent scattering of two-level atoms \cite{Morice1995, Rouabah2014, Corman2017}, we obtain the steady state solutions of Eq. (\ref{sigma13}) by considering a second-order perturbation, which reads
\bea
0=A\tilde\sigma_{13}^{(2)}+\frac{i\Omega_p}{2}-f_{C}\tilde\sigma_{13}^{(1)}.
\eea
From the above and using the result of $\tilde\sigma_{13}^{(1)}$, we have
\bea
\tilde\sigma_{13}^{(2)}=\frac{-i\Omega_p}{2}\left[\frac{1}{A}+\frac{f_{C}}{A^2}+\frac{f_{C}^2}{A^3}\right].\label{multiple}
\eea
Again from Eq. (\ref{MB}), we obtain the output field, 
\bea
\log\left(\frac{\Omega_p(L)}{\Omega_p(0)}\right)&=&\frac{D\Gamma}{4L}\int_0^Ldz\left[\frac{1}{A}+\frac{f_{C}}{A^2}+\frac{f_{C}^2}{A^3}\right],\nonumber\\
&\approx&\frac{D\Gamma}{4}\frac{1}{A-f_{C}},
\eea
where the approximate form can be derived from the first line by using Taylor expansion, $(1-x)^{-1}$ $=$ $\sum_{n=0}x^n$ for $x$ $<$ $1$. 


\begin{thebibliography}{99}
\bibitem{Harris1990} S. E. Harris, J. E. Field, and A. Imamo\ifmmode \breve{g}\else \u{g}\fi{}lu, Phys. Rev. Lett. {\bf 64}, 1107 (1990).
\bibitem{Harris1997} S. E. Harris, Physics Today {\bf 50}, 36 (1997).
\bibitem{Lukin2003} M. D. Lukin, Rev. Mod. Phys. {\bf 75}, 457 (2003).
\bibitem{Fleischhauer2005} M. Fleischhauer, A. Imamoglu, and J. P. Marangos, Rev. Mod. Phys. {\bf 77}, 633 (2005).
\bibitem{Hammerer2010} K. Hammerer, A. S. S\o{}rensen, and E. S. Polzik, Rev. Mod. Phys. {\bf 82}, 1041 (2010).
\bibitem{Hau1999} L. V. Hau, S. E. Harris, Z. Dutton, and C. H. Behroozi, Nature {\bf 397}, 594 (1999).
\bibitem{Schnorrberger2009} U. Schnorrberger, J. D. Thompson, S. Trotzky, R. Pugatch, N. Davidson, S. Kuhr, and I. Bloch, Phys. Rev. Lett. {\bf 103} 033003 (2009).
\bibitem{Fleischhauer2002} M. Fleischhauer and M. D. Lukin, Phys. Rev. A {\bf 65}, 022314 (2002).
\bibitem{Hsiao2018} Y.-F. Hsiao, P.-J. Tsai, H.-S. Chen, S.-X. Lin, C.-C. Hung, C.-H. Lee, Y.-H. Chen, Y.-F. Chen, I. A. Yu, and Y.-C. Chen, Phys. Rev. Lett. {\bf 120}, 183602 (2018).
\bibitem{Wang2019} Y.Wang, J. Li, S. Zhang, K. Su, Y. Zhou, K. Liao, S. Du, H. Yan, and S.-L. Zhu, Nat. Photon. {\bf 13}, 346 (2019).
\bibitem{Saffman2010} M. Saffman, T. G. Waller, and K. Mølmer, Rev. Mod. Phys. {\bf 82}, 2313 (2010).
\bibitem{Pritchard2010} J. D. Pritchard, D. Maxwell, A. Gauguet, K. J. Weatherill, M. P. A. Jones, C. S. Adams, Phys. Rev. Lett. {\bf 105}, 193603 (2010).
\bibitem{Dudin2012} Y. O. Dudin and A. Kuzmich, Science {\bf 336}, 887 (2012).  
\bibitem{Peyronel2012} T. Peyronel, O. Firstenberg, Q. Liang, S. Hofferberth, A. V. Gorshkov, T. Pohl, M. D. Lukin, and V. Vuletic, Nature {\bf 488}, 57 (2012).
\bibitem{Mucke2010} M. M\"{u}cke, E. Figueroa, J. Bochmann, C. Hahn, K. Murr, S. Ritter, C. J. Villas-Boas, and G. Rempe, Nature {\bf 465}, 755 (2010).
\bibitem{Kampschulte2014} T. Kampschulte, W. Alt, S. Manz, M. Martinez-Dorantes, R. Reimann, S. Yoon, D. Meschede, M. Bienert, and G. Morigi, Phys. Rev. A {\bf 89}, 033404 (2014).
\bibitem{Rohlsberger2012} R. R\"{o}hlsberger, H.-C. Wille, K. Schlage, and B. Sahoo, Nature {\bf 482}, 199 (2012).
\bibitem{Hemmer2001} P. R. Hemmer, A. V. Turukhin, M. S. Shahriar, and J. A. Musser, Opt. Lett. {\bf 26}, 361 (2001).
\bibitem{Acosta2013} V. M. Acosta, K. Jensen, C. Santori, D. Budker, and R. G. Beausoleil, Phys. Rev. Lett. {\bf 110}, 213605 (2013).
\bibitem{Serapiglia2000} G. B. Serapiglia, E. Paspalakis, C. Sirtori, K. L. Vodopyanov, and C. C. Phillips, Phys. Rev. Lett. {\bf 84}, 1019 (2000).
\bibitem{Ham1997} B. S. Ham, P. R. Hemmer, and M. S. Shahriar, Opt. Comm. {\bf 144}, 227 (1997).
\bibitem{Turukhin2001} A. V. Turukhin, V. S. Sudarshanam, M. S. Shahriar, J.A. Musser, B. S. Ham, and P. R. Hemmer, Phys. Rev. Lett. {\bf 88}, 023602 (2001).
\bibitem{Longdell2005} J. J. Longdell, E. Fraval, M. J. Sellars, and N. B. Manson, Phys. Rev. Lett. {\bf 95}, 063601 (2005).
\bibitem{Baldit2010} E. Baldit, K. Bencheikh, P. Monnier, S. Briaudeau, J.A. Levenson, V. Crozatier, I. Lorger\'e, F. Bretenaker, J. L. Le Gou\"et, O. Guillot-No\"el, and Ph. Goldner, Phys. Rev. B {\bf 81}, 144303 (2010).
\bibitem{Heinze2013} G. Heinze, C. Hubrich, and T. Halfmann, Phys. Rev. Lett. {\bf 111}, 033601 (2013).
\bibitem {QO:Scully} M. O. Scully and M. S. Zubairy, {\em Quantum Optics} (Cambridge University Press, 1997).
\bibitem{Jiang2009} L. Jiang, H. Pu, W. Zhang, and H. Y. Ling, Phys. Rev. A {\bf 80}, 033606 (2009).
\bibitem{Jen2013} H. H. Jen and Daw-Wei Wang, Phys. Rev. A {\bf 87}, 061802(R) (2013).
\bibitem{Jen2014} H. H. Jen and Daw-Wei Wang, J. Opt. Soc. Am. B {\bf 31}, 2931 (2014).
\bibitem{Lehmberg1970} R. H. Lehmberg, Phys. Rev. A {\bf 2}, 883 (1970).
\bibitem{Dicke1954} R. H. Dicke, Phys. Rev. {\bf 93}, 99 (1954).
\bibitem{Gross1982} M. Gross and S. Haroche, Phys. Rep. {\bf 93} 301 (1982).
\bibitem{Bromley2016} S. L. Bromley, B. Zhu, M. Bishof, X. Zhang, T. Bothwell, J.
Schachenmayer, T. L. Nicholson, R. Kaiser, S. F. Yelin, M. D. Lukin, A. M. Rey, and J. Ye, Nat. Commun. 7:11039 (2016).
\bibitem{Guerin2016} W. Guerin, M. O. Ara\'{u}jo, and R. Kaiser, Phys. Rev. Lett. {\bf 116}, 083601 (2016).
\bibitem{Sonnefraud2010} Y. Sonnefraud, N. Verellen, H. Sobhani, G. A.E. Vandenbosch, V. V. Moshchalkov, P. V. Dorpe, P. Nordlander, and S. A. Maier, ACS Nano {\bf 4}, 1664 (2010).
\bibitem{McGuyer2015} B. H. McGuyer, M. McDonald, G. Z. Iwata, M. G. Tarallo, W. Skomorowski, R. Moszynski, and T. Zelevinsky, Nat. Phys. {\bf 11}, 32 (2015).
\bibitem{Jenkins2017} S. D. Jenkins, J. Ruostekoski, N. Papasimakis, S. Savo, and N. I. Zheludev, Phys. Rev. Lett. {\bf 119}, 053901 (2017).
\bibitem{Rohlsberger2010} R. R\"{o}hlsberger, K. Schlage, B. Sahoo, S. Couet, and R. R\"{u}ffer, Science {\bf 328}, 1248 (2010).
\bibitem{Keaveney2012} J. Keaveney, A. Sargsyan, U. Krohn, I. G. Hughes, D. Sarkisyan, and C. S. Adams, Phys. Rev. Lett. {\bf 108}, 173601 (2012).
\bibitem{Meir2014} Z. Meir, O. Schwartz, E. Shahmoon, D. Oron, and R. Ozeri, Phys. Rev. Lett. {\bf 113}, 193002 (2014).
\bibitem{Pellegrino2014} J. Pellegrino, R. Bourgain, S. Jennewein, Y. R. P. Sortais, A. Browaeys, S. D. Jenkins, and J. Ruostekoski, Phys. Rev. Lett. {\bf 113}, 133602 (2014).
\bibitem{Jennewein2016} S. Jennewein, M. Besbes, N. J. Schilder, S. D. Jenkins, C. Sauvan, J. Ruostekoski, J.-J. Greffet, Y. R. P. Sortais, and A. Browaeys, Phys. Rev. Lett. 116, 233601 (2016).
\bibitem{Jenkins2016} S. D. Jenkins, J. Ruostekoski, J. Javanainen, R. Bourgain, S. Jennewein, Y. R. P. Sortais, and A. Browaeys, Phys. Rev. Lett. 116, 183601 (2016).
\bibitem{Roof2016} S. J. Roof, K. J. Kemp, M. D. Havey, and I. M. Sokolov, Phys. Rev. Lett. {\bf 117}, 073003 (2016).
\bibitem{Oliveira2021} M. H. Oliveira, C. E. Máximo, and C. J. Villas-Boas, Phys. Rev. A {\bf 104}, 063704 (2021). 
\bibitem{Jennewein2018} S. Jennewein, L. Brossard, Y. R. P. Sortais, A. Browaeys, P. Cheinet, J. Robert, and P. Pillet, Phys. Rev. A {\bf 97}, 053816 (2018).
\bibitem{Araujo2016} M. O. Ara\'ujo, I. Kre\ifmmode \check{s}\else \v{s}\fi{}i\ifmmode \acute{c}\else \'{c}\fi{}, R. Kaiser, and W. Guerin, Phys. Rev. Lett. {\bf 117}, 073002 (2016).
\bibitem{Sutherland2016} R. T. Sutherland and F. Robicheaux, Phys. Rev A {\bf 93}, 023407 (2016). 
\bibitem{Chomaz2012} L. Chomaz, L. Corman, T. Yefsah, R. Desbuquois, and J. Dalibard, New J. Phys. {\bf 14}, 055001 (2012).
\bibitem{Tannoudji1992} C. Cohen-Tannoudji, J. Dupont-Rock, and G. Grynberg, {\it Atom-photon interactions: Basic processes and applications} (John Wiley \& Sons, 1992).
\bibitem{Kubo1962} R. Kubo, J. Phys. Soc. Jap. Vol. {\bf 17}, 1100 (1962).
\bibitem{Drummond1987} P. D. Drummond and S. J. Carter, J. Opt. Soc. Am. B {\bf 4}, 1565 (1987).
\bibitem{Sevincli2011} S. Sevin\ifmmode \mbox{\c{c}}\else \c{c}\fi{}li, N. Henkel, C. Ates, and T. Pohl, Phys. Rev. Lett. {\bf 107}, 153001 (2011).
\bibitem{Jen2016_SR} H. H. Jen, M.-S. Chang, and Y.-C. Chen, Phys. Rev. A {\bf 94}, 013803 (2016).
\bibitem{Jen2017_MP} H. H. Jen, Phys. Rev. A {\bf 96}, 023814 (2017).
\bibitem{Robicheaux2020} F. Robicheaux and R. T. Sutherland, Phys. Rev. A {\bf 101}, 013805 (2020). 
\bibitem{Andreoli2021} F. Andreoli, M. J. Gullans, A. A. High, A. Browaeys, and D. E. Chang, Phys. Rev. X {\bf 11}, 011026 (2021). 
\bibitem{Rehler1971} N. E. Rehler and J. H Eberly, Phys. Rev A {\bf 3}, 1735 (1971).
\bibitem{Mazets2007} I. E. Mazets and G. Kurizski, J. Phys. B {\bf 40}, F105 (2007).
\bibitem{Scully2009} M. O. Scully, Phys. Rev. Lett. {\bf 102}, 143601 (2009).
\bibitem{Jen2015} H. H. Jen, Ann. of Phys. (N.Y.) {\bf 360}, 556 (2015).
\bibitem{Lukin1997}  M. D. Lukin, M. Fleischhauer, A. S. Zibrov, H. G. Robinson, V. L. Velichansky, L. Hollberg, and M. O. Scully, Phys. Rev. Lett. {\bf 79}, 2959 (1997).
\bibitem{Wei2020} Y.-C. Wei, B.-H. Wu, Y.-F. Hsiao, P.-J. Tsai, and Y.-C. Chen, Phys. Rev. A {\bf 102}, 063720 (2020). 
\bibitem{Petrosyan2011} D. Petrosyan, J. Otterbach, and M. Fleischhauer, Phys. Rev. Lett. {\bf 107}, 213601 (2011).
\bibitem{Garcia2017} A. Asenjo-Garcia, M. Moreno-Cardoner, A. Albrecht, H. J. Kimble, D. E. Chang, Phys. Rev. X {\bf 7}, 031024 (2017).
\bibitem{Cho2016} Y.-W. Cho, G. T. Campbell, J. L. Everett, J. Bernu, D. B. Higginbottom, M. T. Cao, J. Geng, N. P. Robins, P. K. Lam, and B. C. Buchler, Optica {\bf 3}, 100 (2016). 
\bibitem{Gris2018} P. Vernaz-Gris, K. Huang, M. Cao, A. S. Sheremet, and J. Laurat, Nat. Commun. {\bf 9}, 363 (2018).  
\bibitem{Gardiner2000} C. W. Gardiner and P. Zoller, {\it Quantum Noise: A Handbook of Markovian and Non-Markovian Quantum Stochastic Methods with Applications to Quantum Optics}, 2nd ed. (Springer-Verlag Berlin, 2000). 
\bibitem{Jen2012} H. H. Jen, Phys. Rev. A {\bf 85}, 013835 (2012). 
\bibitem{Hsiao2020} S.-S. Hsiao, K.-T Chen, and I. A. Yu, Opt. Express {\bf 28}, 28414 (2020).
\bibitem{Kim2021} B. Kim, {\it et al.}, Commun. Phys. {\bf 4}, 101 (2021).
\bibitem{Bienias2020} P. Bienias, {\it et al.}, Phys. Rev. Res. {\bf 2}, 033049 (2020).
\bibitem{Tebben2021} A. Tebben, C. Hainaut, A. Salzinger, S. Geier, T. Franz, T. Pohl, M. Gärttner, G. Zürn, and M. Weidemüller, Phys. Rev. A {\bf 103}, 063710 (2021).
\bibitem{Morice1995} O. Morice, Y. Castin, and J. Dalibard, Phys. Rev. A {\bf 51}, 3896 (1995).
\bibitem{Rouabah2014} M.-T. Rouabah, M. Samoylova, R. Bachelard, P. W. Courteille, R. Kaiser, and N. Piovella, JOSA A {\bf 31}, 1031 (2014).
\bibitem{Corman2017} L. Corman, J. L. Ville, R. Saint-Jalm, M. Aidelsburger, T. Bienaim\'e, S. Nascimb\`ene, J. Dalibard, and J. Beugnon, Phys. Rev. A {\bf 96}, 053629 (2017).
\end{thebibliography}
\end{document}